\documentclass[conference]{IEEEtran}
\IEEEoverridecommandlockouts

\usepackage[cmex10]{amsmath}
\usepackage[utf8x]{inputenc}
\usepackage[T1]{fontenc}
\usepackage[american]{babel}
\usepackage{textcomp,amsmath,amssymb,amsfonts,bm,bbm,ucs,cite,mathalfa,upgreek}
\usepackage{balance}
\usepackage{diagbox,booktabs,multirow,longtable}
\usepackage{xcolor, algorithmic,graphicx,subfigure, siunitx}

\newcommand*{\herm}{^{\mathsf{H}}}
\newcommand*{\transp}{^{\mathsf{T}}}

\newcommand{\e}{\mathrm{e}}
\renewcommand{\i}{\mathrm{i}}

\newcommand{\testr}{\mathrel{\underset{\mathcal{H}_{0,r}^{\rm seq}}{\overset{{\mathcal{H}_{1,r}^{\rm seq}}}{\gtrless}}}}
\newcommand{\testt}{\mathrel{\underset{\mathcal{H}_{0,t}^{\rm seq}}{\overset{{\mathcal{H}_{1,t}^{\rm seq}}}{\gtrless}}}}

\DeclareMathOperator*{\argmax}{\arg\;\max}

\begin{document}
\bstctlcite{BSTcontrol}	

\title{STAR-RIS-based Pulse-Doppler Radars}

\author{\IEEEauthorblockN{Emanuele Grossi\IEEEauthorrefmark{1}\IEEEauthorrefmark{2}, Hedieh~Taremizadeh\IEEEauthorrefmark{1}, Luca Venturino\IEEEauthorrefmark{1}\IEEEauthorrefmark{2}}
\IEEEauthorblockA{\IEEEauthorrefmark{1}\textit{Department of Electrical and Information Engineering, University of Cassino and Southern Lazio, 03043 Cassino, Italy}}
\IEEEauthorblockA{\IEEEauthorrefmark{2}\textit{National Inter-University Consortium for Telecommunications, 43124 Parma, Italy}}
\IEEEauthorblockA{E-mails: e.grossi@unicas.it, hedieh.taremizadeh@unicas.it, l.venturino@unicas.it. }
\thanks{This work was supported by the Italian Ministry of Education, University, and Research with the program ``Dipartimenti di Eccellenza 2018--2022'' and by the European Union -- NextGenerationEU -- National Recovery and Resilience Plan, Mission 4, Component 2, Investment 1.1, Call PRIN 2022 D.D. 104 02/02/2022 (Project 202238BJ2R ``CIRCE,'' CUP H53D23000420006).}
\vspace*{-0.2cm}
}

\maketitle

\begin{abstract}
In this study, we consider a pulse-Doppler radar relying on a simultaneously transmitting and reflecting reconfigurable intelligent surface (STAR-RIS) for scanning a given volume; the radar receiver is collocated with the STAR-RIS and aims to detect moving targets and estimate their radial velocity in the presence of clutter. To separate the echoes received from the transmissive and reflective half-spaces, the STAR-RIS superimposes a different slow-time modulation on the pulses redirected in each half-space, while the radar detector employs a decision rule based on a generalized information criterion (GIC). Two scanning policies are introduced, namely, simultaneous and sequential scanning, with different tradeoffs in terms of radial velocity estimation accuracy and complexity of the radar detector.
\end{abstract}
\begin{IEEEkeywords}
Pulse-Doppler radar, scanning radar, target detection, velocity estimation, STAR-RIS, GIC.
\end{IEEEkeywords}

\section{Introduction}
Reconfigurable intelligent surfaces (RISs) are receiving increasing attention in both wireless communication and sensing applications. An RIS is a planar structure consisting of a large number of sub-wavelength size elements (atoms) that are low-cost and reconfigurable in terms of their electromagnetic responses. RISs can improve the spectral efficiency, energy efficiency, security, and reliability of wireless communication networks by controlling to some extend the radio propagation environment~\cite{Di_Renzo_2020,Bjornson-2022,Manzoor-2023}; they have also been considered in radio localization and mapping~\cite{2020-Wymeersch-Localization-and-Mapping}, path planning~\cite{Mu_2021}, and radar target detection~\cite{Buzzi_2021, Buzzi-2022, Taremizadeh-2023}. While transmitting-only or reflecting-only RISs provide half-space coverage, simultaneous transmitting and reflecting RISs (STAR-RISs) allow full-space coverage. The hardware description of a STAR-RIS has been presented in~\cite{Xu_2021}, and, based on the field equivalence principle, a model for the signals transmitted and reflected by each atom has been provided. In~\cite{Schober-STAR-RIS-2022}, three operating protocols have been proposed for a STAR-RIS, namely, energy splitting, mode switching, and time switching. While past studies have mainly focused on using STAR-RISs in communication applications, to the best of the authors' knowledge, no previous work have investigated their specific usage in pulse-Doppler radars.  
\begin{figure}[!t]
 \centerline{\includegraphics[width=0.7\columnwidth]{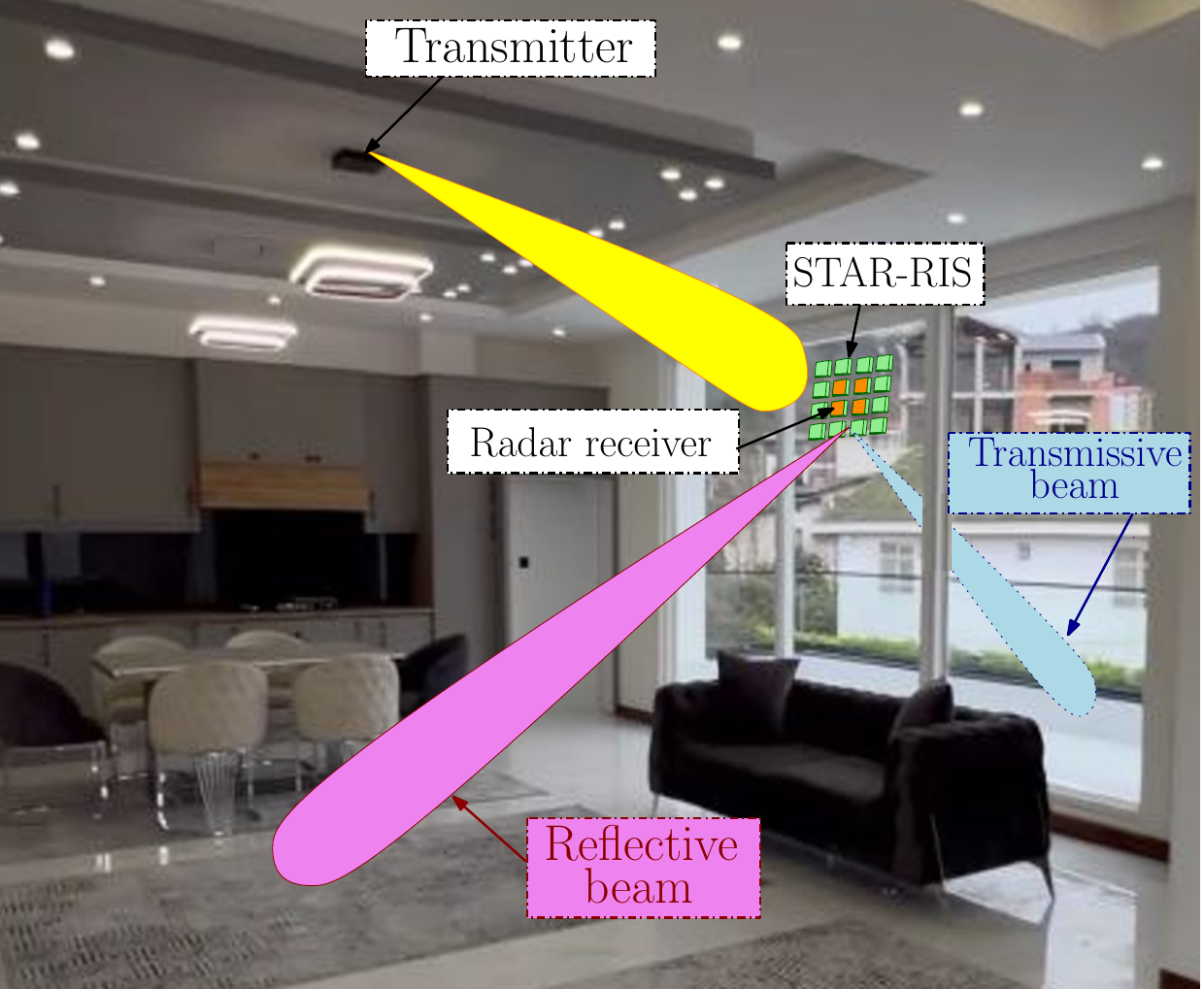}}
	\caption{Graphical description of a STAR-RIS-based pulse-Doppler radar}
	\label{Fig_01}
   \vspace*{-0.25cm}
\end{figure}

In this work, we propose the STAR-RIS-based pulse-Doppler radar architecture in Fig.~\ref{Fig_01}, aiming to detect moving targets on both sides of the STAR-RIS in the presence of clutter. The transmitter illuminates the STAR-RIS that in turn redirects the signal in the transmissive and reflective half-spaces, while the radar receiver elaborates the reverberation from the environment. The receive antennas are collocated with the STAR-RIS and cover both half-spaces, so that a single processing chain implementing passband-to-baseband and analog-to-digital conversion is sufficient. The transmitter may be a dedicated feeder for a stand-alone system or an existing access point which remotely enables the radar sensing at the STAR-RIS. To separate the echoes from the transmissive and reflective half-spaces, we propose to superimpose a slow-time (i.e., from pulse-to-pulse) amplitude/phase modulation on the signals redirected  by the STAR-RIS during a coherent processing interval (CPI); in particular, two practical scanning policies are introduced, referred to as simultaneous and sequential scanning, employing binary phase and on-off amplitude codes, respectively. At the radar receiver, a rule based on a generalized information criterion (GIC)~\cite{Stoica-2004,Grossi-2021} is employed to detect prospective targets on both sides of the STAR-RIS and estimate their radial velocity. Finally, a numerical example is provided to verify the effectiveness of the considered radar architecture and compare the performance-complexity tradeoffs of the proposed scanning policies.

The remainder of this paper is organized as follows. Sec.~\ref{SEC_System_Description} contains the system description.\footnote{Column vectors/matrices are denoted by lower/uppercase boldface letters. The symbols $\i$,  $(\,\cdot\,)^*$, $(\,\cdot\,)\transp$, $(\,\cdot\,)\herm$, $\odot$, and $\otimes$ denote the imaginary unit, conjugate, transpose, conjugate-transpose, the Schur product, and Kronecker product, respectively. $\bm{I}_{M}$ is the $M\times M$ identity matrix. $\text{diag}\{\bm{x}\}$ is a diagonal matrix with the elements of $\bm{x}$ on the main diagonal.} Sec.~\ref{SEC_System_Design} illustrates the design of the STAR-RIS and radar receiver. Sec.~\ref{SEC_Numerical_Analysis} contains  the numerical analysis. Finally, the conclusions are given in Sec.~\ref{SEC_Conclusions}. 

\section{System Description}\label{SEC_System_Description}
Consider the STAR-RIS-based pulse-Doppler radar in Fig.~\ref{Fig_01} operating with a carrier frequency $f_{o}$ and encompassing a transmitter equipped with a directional antenna, a STAR-RIS with $N_{\rm ris}$ tunable atoms, and a radar receiver equipped with $N_{\rm rx}$ antennas. The whole space is divided by the STAR-RIS into two regions, namely, the transmissive and reflective  half-spaces, with the former containing the transmitter. In this study, we make the following assumptions.
\begin{itemize}
\item The atoms of the STAR-RIS and the receive antennas are organized into two collocated uniform rectangular arrays; these arrays are on the $(y,z)$-plane of a Cartesian reference system centered at their center of gravity, with the positive $x$-axis pointing towards the reflective half-space; accordingly, the azimuth angle belongs to $(\pi/2,3\pi/2)$ and $(-\pi/2,\pi/2)$ in the transmissive and reflective half-spaces, respectively.\footnote{Here, the azimuth angle of a point $\bm{p}\in\mathbb{R}^3$ is the angle between the $x$-axis and the orthogonal projection of the vector pointing towards $\bm{p}$ onto the $(x,y)$-plane, which is positive when going from the positive $x$-axis towards the positive $y$-axis, while the elevation angle is the angle between the  vector pointing towards $\bm{p}$ and its orthogonal projection onto the $(x,y)$-plane, which is positive when going towards the positive $z$-axis from the $(x,y)$-plane.}

\item  The baseband signal emitted by the transmitter is a periodic waveform with pulse repetition interval (PRI) $T$ and average power $\mathcal{P}$, namely, $s(t)=\sum_{p=-\infty}^{+\infty}\sqrt{\mathcal{P}T}\psi(t-p T)$, where $\psi(t)$ is a unit-energy pulse with bandwidth $B$ and support $[0,\Delta]$. The transmitter, STAR-RIS, and receiver are perfectly synchronized.

\item The size of both the STAR-RIS and the receive array is much smaller than $c/B$, where $c$ is the speed of light: this is the usual narrowband assumption~\cite{Van-Trees-IV}; also, any mutual coupling among their atoms/antennas is neglected.

\item The radar receiver operates when the receive array (that is collocated with the STAR-RIS) is not illuminated by the transmitter to avoid direct interference. Interestingly, this implies the atoms of the STAR-RIS may be used also for sensing, if equipped with the necessary circuitry~\cite{Zheng-2023,Alexandropoulos-2023}; in this latter case, the controller sets the atoms into the transmitting and/or reflecting mode when the pulses emitted by the transmitter hit the surface and into the sensing mode otherwise~\cite{Zheng-2023,Alexandropoulos-2023}. 

\item The channel between the transmitter and STAR-RIS is known and denoted by $\bm{g} \in \mathbb{C}^{N_{\rm ris}}$.

 \item The inspected region is in the far-field of the STAR-RIS and receive array. We denote by  $\bm{u}_{\rm ris}(\bm{\phi}) \in \mathbb{C}^{N_{\rm ris}}$ and $\bm{u}_{\rm rx}(\bm{\phi}) \in \mathbb{C}^{N_{\rm rx}}$ the corresponding steering vectors towards the direction $\bm{\phi}=[\phi^{\rm az};\phi^{\rm el}]$, respectively, where $\phi^{\rm az}$ and $\phi^{\rm el}$ are the azimuth and elevation angles. 
 \end{itemize}

\subsection{RIS response and power beam-pattern}
The response of the STAR-RIS can be reconfigured at every PRI; in particular, we denote by $\bm{x}_{t}(p)\in\mathbb{C}^{N_{\rm ris}}$ and $\bm{x}_{r}(p)\in\mathbb{C}^{N_{\rm ris}}$ the vectors containing the reflection and transmission coefficients of its atoms during the $p$-th PRI, respectively, with 
\begin{equation}\label{enery_conservation}
	\big|[\bm{x}_{t}(p)]_{n} \big|^2+\big|[\bm{x}_{r}(p)]_{n} \big|^2=1,  
\end{equation}
for $n=1,\ldots,N_{\rm ris}$. Accordingly, the $p$-th pulse redirected by the STAR-RIS towards the direction $\bm{\phi}=[\phi^{\rm az};\phi^{\rm el}]$ is
\begin{equation}\label{eq:radiated_signal}
 \sqrt{\mathcal{P}T G_{\rm ris}(\bm{\phi})}\Big(\bm{u}_{\rm ris}\transp(\bm{\phi})\mathrm{diag}\{\bm{x}(\bm{\phi},p)\}\bm{g}\Big)\psi(t-\delta-p T),
\end{equation}
where $G_{\rm ris}(\bm{\phi})$ is the element gain of the STAR-RIS, $\delta$ is the propagation delay between the transmitter and STAR-RIS,  and
\begin{equation}\label{eq:xr_xt}
\bm{x}(\bm{\phi},p)=\begin{cases}		
		\bm{x}_{t}(p), & \phi^{\rm az}\in(\pi/2,3\pi/2),\\
\bm{x}_{r}(p), & \phi^{\rm az}\in(-\pi/2,\pi/2).
\end{cases}	
\end{equation}

It is seen from~\eqref{eq:radiated_signal} that $\bm{x}_{t}(p)$ and $\bm{x}_{r}(p)$ determine the power beampattern of the STAR-RIS during the $p$-th PRI, namely,
\begin{equation}\label{beampattern}
\textrm{BP}_{\rm ris}(\bm{\phi},p)=G_{\rm ris}(\bm{\phi})\underbrace{\Big|\bm{u}_{\rm ris}\transp(\bm{\phi})\mathrm{diag}\{\bm{x}(\bm{\phi},p)\}\bm{g}\Big|^2}_{\textrm{GF}_{\rm ris}(\bm{\phi},p)}, 
\end{equation}
where $\textrm{GF}_{\rm ris}(\bm{\phi},p)$ is the array gain factor; hence, these vectors can be chosen to focus the STAR-RIS towards some desired angular directions~\cite{Taremizadeh-2023} and, possibly, to introduce an slow-time modulation in the redirected signals~\cite{Qiang-2021}. Let $\mathcal{F}_{t}$ and $\mathcal{F}_{r}$ be the sets of angular directions to be monitored in the transmissive and reflective half-spaces, respectively. Over a CPI of length $PT$, the radar aims to inspect two angular directions, say $\bm{\phi}_{t}=[\phi_{t}^{\rm az};\phi_{t}^{\rm el}]\in\mathcal{F}_{t}$ and $\bm{\phi}_{r}=[\phi_{r}^{\rm az};\phi_{r}^{\rm el}]\in\mathcal{F}_{r}$, one  in the transmissive  half-space and one in the reflective half-space.

\subsection{Radar received signal} 
For illustration, consider the CPI spanning the time segment $[0,PT)$, and denote by $\bm{b}(t)\in \mathbb{C}^{N_{\rm rx}}$ the baseband continuous-time signal collected by the receiver; also, denote by $\mathcal{T}$ the set of delays to be inspected by the radar, each one corresponding to a different range cell. For a given $\tau\in\mathcal{T} $, matched-filter pulse compression provides the following data samples
\begin{multline}
	\bm{y}(p)=\int_{\mathbb{R}} \bm{b}(t) \psi^{*}\big(t-\delta-\tau-(p-1)T\big)dt\;\in \mathbb{C}^{N_{\rm rx}},
\end{multline} 
for $p=1,\ldots,P$. The tuples $(\bm{\phi}_{t},\tau)$ and $(\bm{\phi}_{r},\tau)$ specify the pair of resolution cells under inspection: we refer to them as the transmissive and reflective cells, respectively. Upon assuming that at most one target is present in each resolution cell,  $\bm{y}(p)$ can be expanded as 
\begin{align}
	\bm{y}(p)&=\alpha_{t}\e^{\i 2\pi \nu_{t}T (p-1)}\Big(\bm{u}_{\rm ris}\transp(\bm{\phi}_t)\mathrm{diag}\{\bm{x}_{t}(p)\}\bm{g}\Big) \bm{u}_{\rm rx}(\bm{\phi}_t)\notag \\
	&\quad +\alpha_{r}\e^{\i 2\pi \nu_{r}T (p-1)}\Big(\bm{u}_{\rm ris}\transp(\bm{\phi}_r)\mathrm{diag}\{\bm{x}_{r}(p)\}\bm{g}\Big) \bm{u}_{\rm rx}(\bm{\phi}_r)\notag \\
	&\quad + \bm{z}(p) \in \mathbb{C}^{N_{\rm rx}},\label{eq-observation-1}
\end{align}
where $\alpha_{t}$ and $\nu_{t}$ are the complex amplitude and Doppler shift of a prospective target in the transmissive cell, $\alpha_{r}$ and $\nu_{r}$ are the complex amplitude and Doppler shift of a prospective target in the reflective cell,  and $\bm{z}(p)$ is the additive disturbance, including both clutter and noise (more on this in Sec.~\ref{SEC:Disturbance model}). Here, $\alpha_{t}$  and $\alpha_{r}$ account for the transmitted power $\mathcal{P}$, the two-way path-loss from the STAR-RIS to the target, and the target radar cross-section; clearly, $\alpha_{t}=0$ and $\alpha_{r}=0$ if no target is present in the transmissive and reflective cells, respectively. 

The observations in~\eqref{eq-observation-1} are collected into the vector $\bm{y}=\big[\bm{y}(1);\cdots;\bm{y}(P)\big]$; in particular, upon defining  $\bm{x}_{t}=[\bm{x}_{t}(1);\cdots;\bm{x}_{t}(P)]\in \mathbb{C}^{PN_{\rm ris}}$, $\bm{x}_{r}=[\bm{x}_{r}(1);\cdots;\bm{x}_{r}(P)]\in \mathbb{C}^{PN_{\rm ris}}$, $\bm{d}(\nu)=[1;\e^{\i 2\pi T \nu };\cdots;\e^{\i 2\pi \nu  T(P-1)}]\in \mathbb{C}^{P}$, and $\bm{G}(\phi)=\bm{I}_{P}\otimes \left(\bm{u}_{\rm ris}\transp(\phi)\mathrm{diag}\{\bm{g}\}\right)\in \mathbb{C}^{P\times PN_{\rm ris}}$,
we have
\begin{equation}
	\bm{y}=\alpha_{t} \bm{h}(\bm{x}_{t}, \bm{\phi}_{t},\nu_{t})+\alpha_{r} \bm{h}(\bm{x}_{r}, \bm{\phi}_{r},\nu_{r})+\bm{z} \in \mathbb{C}^{PN_{\rm rx}},	
\end{equation}
where  $\bm{z}=\big[\bm{z}(1);\cdots;\bm{z}(P)\big]$ and 
\begin{equation}\label{h_target}
	\bm{h}(\bm{x}, \bm{\phi},\nu)=\big(\bm{d}(\nu) \odot \bm{G}(\bm{\phi})\bm{x}\big)
	\otimes \bm{u}_{\rm rx}(\bm{\phi})\in \mathbb{C}^{PN_{\rm rx}}	
\end{equation}  
is the \emph{space-time} steering vector of a target with angular direction $\bm{\phi}$ and Doppler shift $\nu$ when the STAR-RIS response vector in its half-space is $\bm{x}$. Notice that $\bm{u}_{\rm rx}(\bm{\phi})$ and $\bm{d}(\nu) \odot \bm{G}(\bm{\phi})\bm{x}$ are the \emph{spatial} and \emph{temporal} steering vectors, respectively, and that a pulse-to-pulse   variation of the STAR-RIS response induces a slow-time modulation in the temporal steering vector: this latter property will be exploited in Sec.~\ref{SEC_scanning_policies} to devise suitable scanning policies.

\subsection{Disturbance model}\label{SEC:Disturbance model}
Assuming $K_{t}$ and $K_{r}$ clutter components in the reflective and transmissive half-spaces, respectively, we have
\begin{equation}
	\bm{z}\!=\!\bm{z}_{n} +\sum_{k=1}^{K_{t}} \alpha_{t,k} \bm{h}(\bm{x}_{t}, \bm{\phi}_{t,k},\nu_{t,k})+\sum_{k=1}^{K_{r}} \alpha_{r,k} \bm{h}(\bm{x}_{r}, \bm{\phi}_{r,k},\nu_{r,k}),
\end{equation}
where $\bm{z}_{n}$ is the additive noise, $\alpha_{t,k}$, $\nu_{t,k}$, and $\bm{\phi}_{t,k}=[\phi_{t,k}^{\rm az};\phi_{t,k}^{\rm el}]$  are the complex amplitude, Doppler shift, and direction of the $k$-th clutter component in the transmissive half-space, respectively, and $\alpha_{r,k}$, $\nu_{r,k}$, and $\bm{\phi}_{r,k}=[\phi_{r,k}^{\rm az};\phi_{r,k}^{\rm el}]$  are the complex amplitude, Doppler shift, and direction of the $k$-th clutter component in the reflective half-space, respectively. 

Hereafter, we model $\bm{z}_{n}$ as a circularly-symmetric Gaussian random vector with covariance matrix $\sigma^{2}_{n}\bm{I}_{PN_{\rm rx}}$ and $\alpha_{t,k}$ and $\alpha_{r,k}$ as independent circularly-symmetric Gaussian random variables with variance $\sigma^{2}_{t,k}$ and $\sigma^{2}_{r,k}$, respectively. Accordingly, the disturbance covariance matrix is
\begin{align}
	\bm{C}&=\sigma^{2}_{n}\bm{I}_{PN_{\rm rx}}+\sum_{k=1}^{K_{t}} \sigma^{2}_{t,k} \bm{h}(\bm{x}_{t}, \bm{\phi}_{t,k},\nu_{t,k})\bm{h}\herm(\bm{x}_{t}, \bm{\phi}_{t,k},\nu_{t,k})\notag \\
	&\quad +\sum_{k=1}^{K_{r}} \sigma^{2}_{r,k} \bm{h}(\bm{x}_{r}, \bm{\phi}_{r,k},\nu_{r,k}) \bm{h}\herm(\bm{x}_{r}, \bm{\phi}_{r,k},\nu_{r,k}).
\end{align}
In the following, $\bm{C}$  is assumed known: in practice, an estimate can be obtained by resorting to parametric or nonparametric algorithms, possibly aided by secondary data and/or some prior knowledge of the surrounding environment~\cite{Capraro-2006,Guerci-2006,Stoica-2011,Zhu-2011}.

\section{System design}\label{SEC_System_Design}

\subsection{Proposed scanning policies}\label{SEC_scanning_policies}
The echoes from the two half-spaces must be separable by the radar detector. Upon looking at~\eqref{h_target}, it is seen that the spatial steering vector alone cannot ensure such separability, since $\bm{u}_{\rm rx}([\phi^{\rm az};\phi^{\rm el}])\propto\bm{u}_{\rm rx}([\pi-\phi^{\rm az};\phi^{\rm el}])$ for $\phi^{\rm az}\in(-\pi/2,\pi/2)$: this is a consequence of the fact that a single receive array simultaneously covers both half-spaces. This limitation can be overcome by exploiting the  temporal steering vector $\bm{d}(\nu) \odot \bm{G}(\bm{\phi})\bm{x}$, which can be controlled by the STAR-RIS: the basic idea is that the STAR-RIS can superimpose a different slow-time modulation on the pulses redirected in each half-space, while focusing towards the desired angular directions. 

To illustrate this idea, assume that $\bm{x}_{t}=\bm{c}_{t} \otimes \bar{\bm{x}}_{t}$ and  $\bm{x}_{r}=\bm{c}_{r} \otimes \bar{\bm{x}}_{r}$ , where $\bar{\bm{x}}_{t}$ and $\bar{\bm{x}}_{r}$ are $N_{\rm ris}$--dimensional vectors with unit-modulus entries, while  $\bm{c}_{t}$ and $\bm{c}_{r}$ are $P$--dimensional vectors with $\big|[\bm{c}_{t}]_{p}\big|^2+\big|[\bm{c}_{r}]_{p}\big|^2=1$ for $p=1,\ldots,P$, so that the constraint in~\eqref{enery_conservation} is satisfied; accordingly, we have
\begin{subequations}
\begin{align}	
 \bm{h}(\bm{x}_{t}, \bm{\phi},\nu)&=\bm{u}_{\rm ris}\transp( \bm{\phi})\mathrm{diag}\{\bar{\bm{x}}_{t}\}\bm{g}\Big[\big(\bm{d}(\nu) \odot\bm{c}_{t}\big)
\otimes \bm{u}_{\rm rx}(\bm{\phi})\Big],\\
\bm{h}(\bm{x}_{r}, \bm{\phi},\nu)&=\bm{u}_{\rm ris}\transp( \bm{\phi})\mathrm{diag}\{\bar{\bm{x}}_{r}\}\bm{g}\Big[\big(\bm{d}(\nu) \odot\bm{c}_{r}\big)
\otimes \bm{u}_{\rm rx}(\bm{\phi})\Big].
\end{align}  \label{h_target_1}%
\end{subequations}
The vectors $\bar{\bm{x}}_{t}$ and $\bar{\bm{x}}_{r}$ control (up to a scaling factor) the array gain factor of the  STAR-RIS in the transmissive and reflective half-spaces, respectively. Different criteria can be adopted to synthesize a desired power beampattern~\cite{Buzzi_2021, Buzzi-2022, Grossi-2023-Beampattern}; for illustration, $\bar{\bm{x}}_{t}$ and $\bar{\bm{x}}_{r}$ are chosen here to maximize the array gain factor towards $\bm{\phi}_{t}$ and $\bm{\phi}_{r}$, respectively, whereby
\begin{subequations}
\begin{align}  
[\bar{\bm{x}}_{t}]_{n}&=\e^{-\i \big(\angle[\bm{g}]_{n}+\angle[\bm{u}_{\rm rx}(\bm{\phi}_{t})]_{n}\big)}, \label{STAR-RIS-spatial_response-t}\\
[\bar{\bm{x}}_{r}]_{n}&=\e^{-\i \big(\angle[\bm{g}]_{n}+\angle[\bm{u}_{\rm rx}(\bm{\phi}_{r})]_{n}\big)}.\label{STAR-RIS-spatial_response-r}
\end{align}\label{STAR-RIS-spatial_response}%
\end{subequations}
Instead, the vectors $\bm{c}_{t}$ and $\bm{c}_{r}$ control the slow-time modulation superimposed on the signals redirected in the transmissive and reflective half-spaces, respectively. In order to separate the echoes from these half-spaces, we propose to use orthogonal code sequences, i.e., $\bm{c}_{t}\herm\bm{c}_{r}=0$. In particular, upon assuming $P$ even, we consider two practical scanning rules.

\noindent \emph{1)~Simultaneous~scanning:}~In this case,  $[\bm{c}_{t}]_{p}=(-1)^n/ \sqrt{2}$ and $[\bm{c}_{r}]_{p}=1/ \sqrt{2}$, for $p=1,\ldots,P$, whereby the STAR-RIS simultaneously illuminates both half-spaces during the entire CPI by equally splitting the incoming energy~\cite{Schober-STAR-RIS-2022}; the superimposed binary phase codes implement here a form of code-division multiple-access.

\noindent \emph{2)~Simultaneous~scanning:}~In this case, $[\bm{c}_{t}]_{p}=1$ and $[\bm{c}_{r}]_{p}=0$ for $p=1,\ldots,P/2$ and $[\bm{c}_{t}]_{p}=0$ and $[\bm{c}_{r}]_{p}=1$ for $p=P/2+1,\ldots,P$, whereby the CPI is partitioned into two sub-intervals of duration $P/2$, and the STAR-RIS illuminates only one half-space in each sub-interval~\cite{Schober-STAR-RIS-2022}; the adopted on-off codes implement here a form of time-division multiple-access.

\subsection{Proposed GIC-based detector}\label{SEC_GIC-based detector}
The radar detector is faced  with a composite hypothesis testing problem  with four hypotheses: under the hypothesis $\mathcal{H}_{0}$, no target is present in both resolution cells; under $\mathcal{H}_{1,t}$ ($\mathcal{H}_{1,r}$), a target is present only in the transmissive (reflective) cell; under $\mathcal{H}_{2}$, a target is present in both resolution cells. The amplitude and Doppler shift of the prospective targets are unknown.  To tackle this problem, we resort to a generalized information criterion~\cite{Stoica-2004,Grossi-2021}, whereby the selected hypothesis is
\begin{equation}\label{GIC-rule}
	\hat{\mathcal{H}}=\argmax_{\mathcal{L}\in\{\mathcal{H}_{0},\mathcal{H}_{1,t},\mathcal{H}_{1,r},\mathcal{H}_{2}\}} \mu(\mathcal{L}),
\end{equation} 
where the objective function is defined as
\begin{equation}
\mu(\mathcal{L})\!=\!	\begin{cases}
		\ln f_{\mathcal{H}_{0}}(\bm{y}), & \text{if } \mathcal{L}=\mathcal{H}_{0},\\
		\displaystyle \max_{\alpha_t\in \mathbb{C}, \nu_{t}\in\mathcal{V}}\ln f_{\mathcal{H}_{1,t}}(\bm{y};\alpha_t,\nu_t)-\eta, & \text{if } \mathcal{L}=\mathcal{H}_{1,t},\\
		\displaystyle\max_{\alpha_r\in \mathbb{C}, \nu_{r}\in\mathcal{V}} \ln f_{\mathcal{H}_{1,r}}(\bm{y};\alpha_r,\nu_r)-\eta, & \text{if } \mathcal{L}=\mathcal{H}_{1,r},\\		
		\displaystyle\max_{\bm{\alpha}\in \mathbb{C}^2, \bm{\nu}\in\mathcal{V}^2} \ln f_{\mathcal{H}_{2}}(\bm{y};\bm{\alpha},\bm{\nu})-2\eta, & \text{if } \mathcal{L}=\mathcal{H}_{2},\\
	\end{cases}
\end{equation}
$f_{\mathcal{L}}(\bm{y};\cdot)$ is the likelihood function under $\mathcal{L}$, $\bm{\alpha}=[\alpha_t;\alpha_r]$, $\bm{\nu}=[\nu_{t};\nu_{r}]$, $\mathcal{V}$ is Doppler search interval, and $\eta$ is a penalty factor that can be set to control the false alarm rate under $\mathcal{H}_{0}$. 

Let $\bm{x}=[\bm{x}_{t};\bm{x}_{r}]$, $\bm{\phi}=[\bm{\phi}_{t};\bm{\phi}_{r}]$, and $\bm{H}(\bm{x}, \bm{\phi},\bm{\nu})=[\bm{h}(\bm{x}_{t}, \bm{\phi}_{t},\nu_{t}),\bm{h}(\bm{x}_{r}, \bm{\phi}_{r},\nu_{r})]$; also, let \begin{subequations}
\begin{align}
\bm{\pi}_{t}(\nu_{t})&=\bm{C}^{-1}\bm{h}(\bm{x}_{t}, \bm{\phi}_{t},\nu_{t})/\|\bm{C}^{-\frac{1}{2}}\bm{h}(\bm{x}_{t}, \bm{\phi}_{r},\nu_{t})\|,\\
\bm{\pi}_{r}(\nu_{r})&=\bm{C}^{-1}\bm{h}(\bm{x}_{r}, \bm{\phi}_{r},\nu_{r})/\|\bm{C}^{-\frac{1}{2}}\bm{h}(\bm{x}_{r}, \bm{\phi}_{r},\nu_{r})\|,\\
\bm{\Pi}(\bm{\nu})&=\bm{C}^{-1}\bm{H}(\bm{x}, \bm{\phi},\bm{\nu})\notag \\&\quad \times \big(\bm{H}\herm(\bm{x}, \bm{\phi},\bm{\nu}) \bm{C}^{-1}\bm{H}(\bm{x}, \bm{\phi},\bm{\nu})\big)^{-1/2}.
\end{align}
\end{subequations}
Upon exploiting the fact that the disturbance is Gaussian and after some elaborations, the rule in~\eqref{GIC-rule} can be recast as~\cite{Grossi-2021}
\begin{equation}\label{GIC-rule-2}
	\hat{\mathcal{H}}=\argmax_{\mathcal{L}\in\{\mathcal{H}_{0},\mathcal{H}_{1,t},\mathcal{H}_{1,r},\mathcal{H}_{2}\}} \tilde{\mu}(\mathcal{L}),
\end{equation} 
where the objective function is defined as
\begin{equation}\label{GIC-rule-2-obj}
	\tilde{\mu}(\mathcal{L})\!=\!	\begin{cases}
	0, & \text{if } \mathcal{L}=\mathcal{H}_{0},\\  
		\displaystyle \max_{\nu_{t}\in\mathcal{V}} 
  \left\| \bm{\pi}_{t}\herm(\nu_{t}) \bm{y}\right\|^2
		-\eta, & \text{if } \mathcal{L}=\mathcal{H}_{1,t},\\
	\displaystyle\max_{\nu_{r}\in\mathcal{V}} 
		\left\| \bm{\pi}_{r}\herm(\nu_{r}) \bm{y}\right\|^2
		-\eta, & \text{if } \mathcal{L}=\mathcal{H}_{1,r},\\
	\displaystyle\max_{ \bm{\nu}\in\mathcal{V}^2} \left\| \bm{\Pi}\herm(\bm{\nu}) \bm{y}\right\|^2-2\eta, & \text{if } \mathcal{L}=\mathcal{H}_{2}.\\
	\end{cases}
\end{equation}
When a target is declared in a given resolution cell, an estimate of its Doppler shift (and therefore of its radial velocity) is recovered from the corresponding argument maximizing the objective function in~\eqref{GIC-rule-2-obj} under $\hat{\mathcal{H}}$.

\subsubsection{Sequential scanning}
Let $\bm{y}_{t}\in\mathbb{C}^{PN_{\rm ris}/2}$ and $\bm{y}_{r}\in\mathbb{C}^{PN_{\rm ris}/2}$ be the vectors containing the first and the last half entries of $\bm{y}$, i.e.,  $\bm{y}=[\bm{y}_{t};\bm{y}_{r}]$. When using a sequential scanning, $\bm{y}_{t}$ and $\bm{y}_{r}$ only contain echoes originated from the transmissive and reflective half-spaces, respectively. Upon exploiting this property, the implementation of~\eqref{GIC-rule-2} can be simplified; it particular, two independent binary tests can be  run in each half-space, namely,
\begin{equation}
  \max_{\nu_{t}\in\mathcal{V}} \left\| \overline{\bm{\pi}}_{t}\herm(\nu_{t}) \bm{y}_{t}\right\|^2\testt \eta, \;\;\; \max_{\nu_{r}\in\mathcal{V}} \left\| \underline{\bm{\pi}}_{r}\herm(\nu_{r}) \bm{y}_{r}\right\|^2\testr \eta, 
\end{equation}
where $\overline{\bm{\pi}}_{t}\in\mathbb{C}^{PN_{\rm ris}/2}$ contains the first half entries of $\bm{\pi}_{t}$, while $\underline{\bm{\pi}}_{r}\in\mathbb{C}^{PN_{\rm ris}/2}$ the second half entries of $\bm{\pi}_{r}$; then, a decision is taken as follows: $\mathcal{H}_{0}$ is declared if $\mathcal{H}_{0,t}^{\rm seq}$ and $\mathcal{H}_{0,r}^{\rm seq}$ are true; $\mathcal{H}_{1,t}$ is declared if $\mathcal{H}_{1,t}^{\rm seq}$ and $\mathcal{H}_{0,r}^{\rm seq}$ are true; $\mathcal{H}_{1,r}$ is declared if $\mathcal{H}_{0,t}^{\rm seq}$ and $\mathcal{H}_{1,r}^{\rm seq}$ are true; $\mathcal{H}_{2}$ is declared if $\mathcal{H}_{1,t}^{\rm seq}$ and $\mathcal{H}_{1,r}^{\rm seq}$ are true. As compared to~\eqref{GIC-rule-2}, not only the joint Doppler search under $\mathcal{H}_{2}$ is avoided, but also the involved vectors and matrices have
smaller size: for example, it can be shown that $\overline{\bm{\pi}}_{t}(\nu_{t})=\overline{\bm{C}}^{(-1)}\overline{\bm{h}}(\bm{x}_{t}, \bm{\phi}_{t},\nu_{t})/\|\overline{\bm{C}}^{(-1/2)}\overline{\bm{h}}(\bm{x}_{t}, \bm{\phi}_{t},\nu_{t})\|$, 
where $\overline{\bm{h}}(\bm{x}_{t}, \bm{\phi}_{t},\nu_{t})$ contains the first half entries of $\bm{h}(\bm{x}_{t}, \bm{\phi}_{t},\nu_{t})$ and $\overline{\bm{C}}$ is the submatrix of $\bm{C}$ with the first half rows and columns.

\section{Numerical results}\label{SEC_Numerical_Analysis}

We consider a system employing a carrier frequency $f_{o}=28$~GHz, a bandwidth $B=50$~MHz, a STAR-RIS and a receive array with $16$ elements along the $y$-axis and $8$ along the $z$-axis (whereby $N_{\rm ris}=N_{\rm rx}=128$), rectangular probing pulses,  $T=0.5$~ms, and $P=8,16,32$. Notice that the unambiguous Doppler interval is here $(-\nu_{\max},\nu_{\max})$, with $\nu_{\max}=1/(2T)=1$~kHz; this latter value corresponds to a radial velocity of $c\nu_{\max}/(2f_{o})\simeq 5.3$~m/s~\cite{book-Richards}, which may be sufficient for low-mobility applications. Also, for the same CPI, simultaneous and sequential scanning presents a different Doppler resolution of $1/(PT)$ and $2/(PT)$, since they elaborate $P$ and $P/2$ consecutive pulses from each resolution cell under inspection, receptively~\cite{book-Richards}; if $P=16$, the Doppler resolution for simultaneous scanning  is $125$~Hz, corresponding to a radial velocity of about $0.7$~m/s.

\begin{figure}[t]
\centerline{\includegraphics[width=0.7\columnwidth]{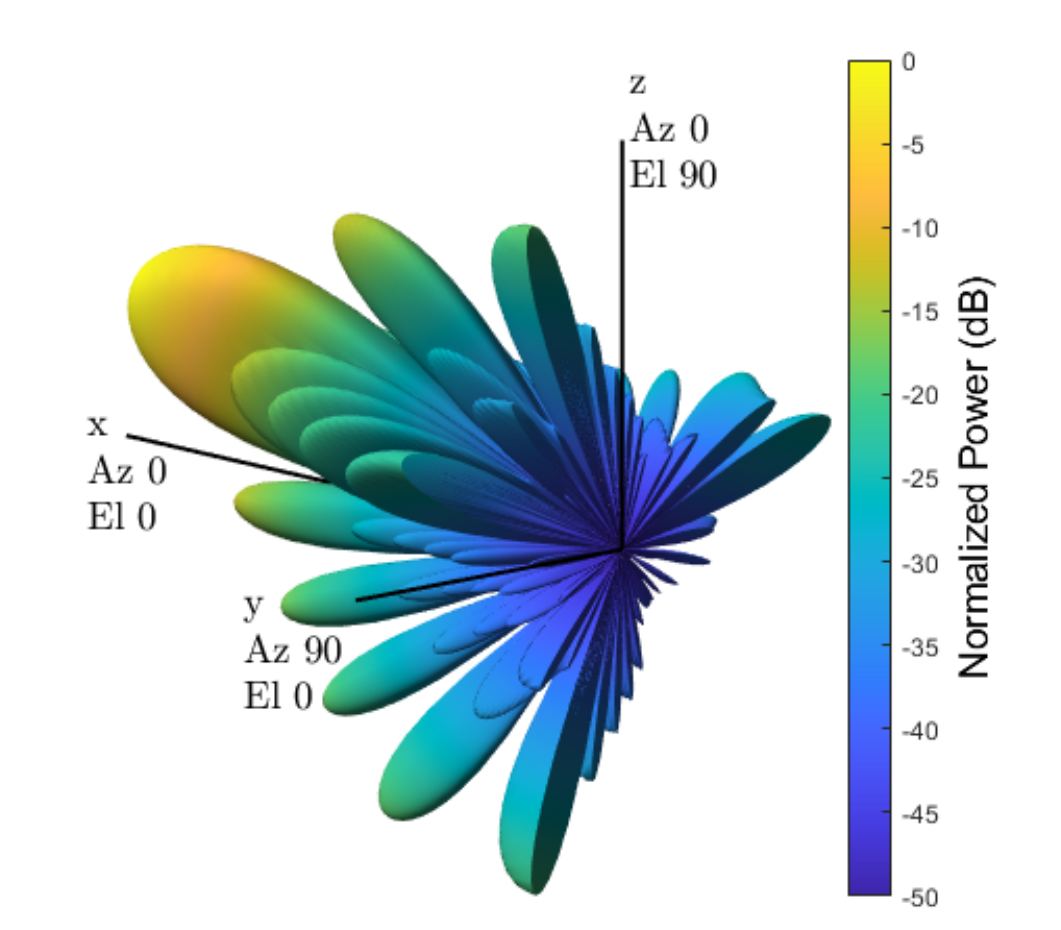}}
 \caption{Normalized array gain factor of the STAR-RIS in the reflective half-space, when the design in~\eqref{STAR-RIS-spatial_response} is employed and $\bm{\phi}_r=[22^{\circ};22^{\circ}]$.}
\label{Fig_03}\vspace{-0.1cm}
\end{figure}

The entries of channel from the transmitter to STAR-RIS are drawn from a complex Gaussian distribution. As to the targets, their directions and Doppler shifts are randomly generated, with  $\bm{\phi}_t \in(155^{\circ},160^{\circ})\times (20^{\circ},25^{\circ})$, $\bm{\phi}_r \in(20^{\circ},25^{\circ})\times (20^{\circ},25^{\circ})$,  and $\nu_{t},\nu_{r}\in(500,1000)$~Hz; also, $\alpha_{t}$ and $\alpha_{r}$ are drawn from a circularly-symmetric Gaussian distribution with variance $\sigma^{2}_{t}$ and $\sigma^{2}_{r}$, respectively, corresponding to a Swerling~I fluctuation model~\cite{book-Richards}. Both targets have the same signal-to-noise ratio per pulse, defined as $\mathrm{SNR}_{\rm p}=\sigma_{t}^2 \|\bm{h}(\bm{x}_{t}, \bm{\phi}_{t},\nu_{t})\|^2/(P\sigma_{n}^2)=
\sigma_{r}^2\|\bm{h}(\bm{x}_{r}, \bm{\phi}_{r},\nu_{r})\|^2/(P\sigma_{n}^2)$. As to clutter components, $K_{t}=K_{r}=10$, and their directions and Doppler shifts are randomly generated, with  $\bm{\phi}_{t,k} \in(200^{\circ},220^{\circ})\times (-40^{\circ},-20^{\circ})$, $\bm{\phi}_{r,q} \in(-40^{\circ},-20^{\circ})\times (-40^{\circ},-20^{\circ})$, and $\nu_{r,k},\nu_{t,q}\in (-125,125)$~Hz, for $k=1,\ldots, K_{t}$ and $q=1,\ldots,K_{r}$; also, they all have a clutter-to-noise ratio per pulse of $20$~dB, defined  as $\mathrm{CNR}_{\rm p}=\sigma_{t,k}^2\|\bm{h}(\bm{x}_{t}, \bm{\phi}_{t,k},\nu_{t,k})\|^2/(P\sigma_{n}^2)=
\sigma_{r,q}^2\|\bm{h}(\bm{x}_{r}, \bm{\phi}_{r,q},\nu_{r,q})\|^2/(P\sigma_{n}^2)$.
The STAR-RIS response is set according to~\eqref{STAR-RIS-spatial_response}; for example, Fig.~\ref{Fig_03} reports the resulting normalized array gain factor in the reflective half-space with $\bm{\phi}_r=[22^{\circ};22^{\circ}]$. Finally, $\eta$ is chosen to have an average number of false alarms per CPI under $\mathcal H_0$ equal to $10^{-3}$.

\begin{figure}[t]%
\centerline{\includegraphics[width=1.1\columnwidth]{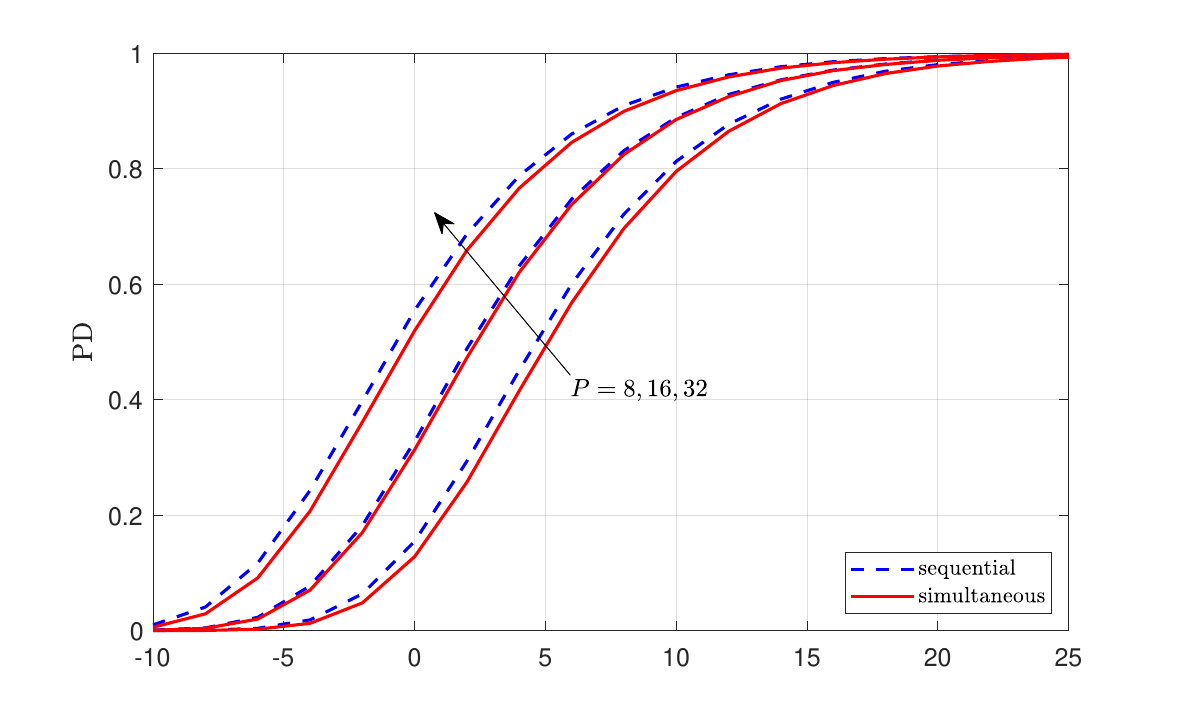}} 
\vspace*{-0.2cm}
\centerline{\includegraphics[width=1.1\columnwidth]{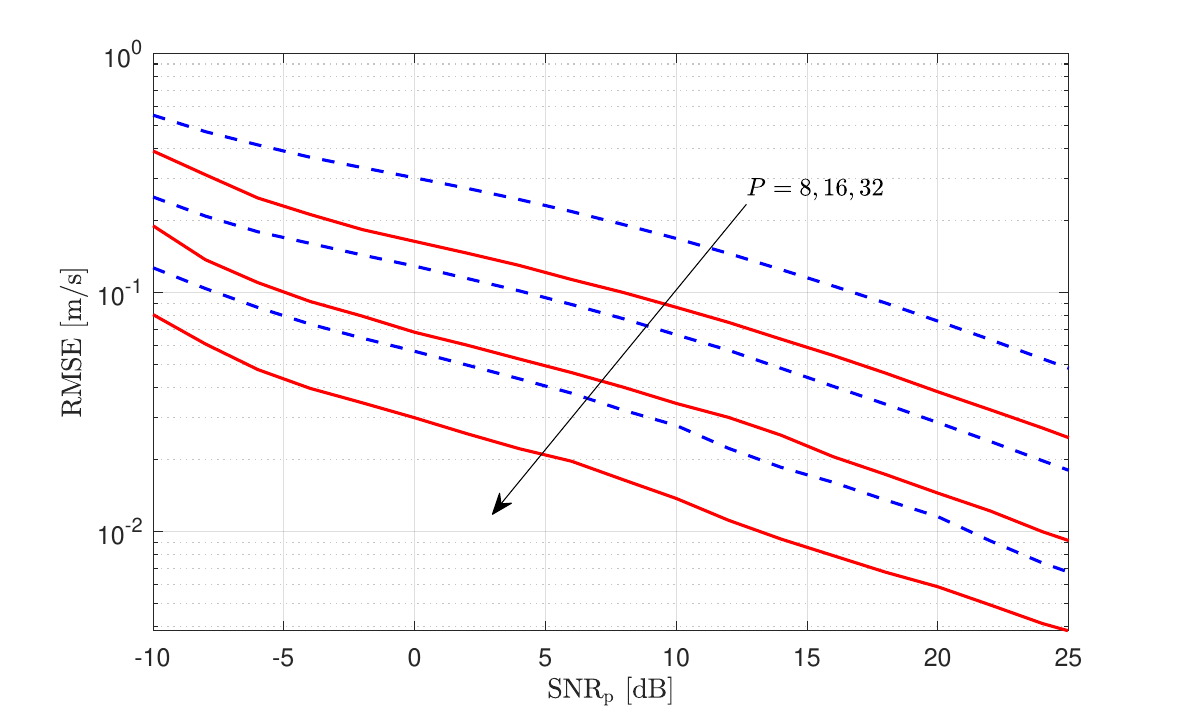}}
\caption{PD and  RMSE in the radial velocity  estimation versus $\mathrm{SNR}_{\rm p}$.} 
\label{fig_PD_RMSE}\vspace{-0.1cm}
\end{figure}

Fig.~\ref{fig_PD_RMSE} reports the probability of declaring $\mathcal{H}_{2}$ when $\mathcal{H}_{2}$ is true (shortly, $\mathrm{PD}$) and the corresponding root mean square error (RMSE) in the estimation of the target radial velocity (averaged over both targets) versus the $\mathrm{SNR}_{\rm p}$, for $P=8,16,32$. Both scanning policies substantially present the same $\mathrm{PD}$ for the same value of $P$, as they redirect the same amount of energy in the two half-spaces in a given CPI. Instead, simultaneous scanning provides a lower RMSE than sequential scanning, since more pulses are elaborated from each resolution cell during the same CPI: this comes at price of a larger implementation complexity of the radar detector, as described in~Sec.~\ref{SEC_GIC-based detector}. Finally, doubling $P$ results into an energy integration gain of $3$~dB for both scanning policies, as evident from the shift of the PD curves, and into a smaller Doppler resolution; when looking at the RMSE curves, both these effects help improving the velocity estimation.  

\section{Conclusions}\label{SEC_Conclusions}
In this paper, we have introduced a STAR-RIS-based pulse-Doppler radar with sensing at the STAR-RIS location. After introducing a convenient signal model, we have proposed two scanning policies to ensure separability of the echoes received from the transmissive  and reflective half-spaces, and we have devised a decision rule based on a generalized information criterion for joint target detection and radial velocity estimation. The analysis indicates that simultaneous scanning provides a better radial velocity estimation than sequential scanning, at the price of a larger complexity of the radar detector. 


\end{document}